\documentclass[prb,showpacs,twocolumn]{revtex4-1}
\usepackage{graphicx}
\usepackage{dcolumn}
\usepackage{bm}
\usepackage{epstopdf}
\usepackage{epsfig}
\usepackage{hyperref}

\begin{document}
\title{Electronic structure and polaronic charge distributions of Fe vacancy clusters in Fe$_{1-x}$O}

\author{I.~Bernal-Villamil, S.~Gallego} 
\email{sgallego@icmm.csic.es}
\affiliation{
Instituto de Ciencia de Materiales de Madrid, Consejo Superior de
Investigaciones Cient{\'{\i}}ficas, Cantoblanco, 28049 Madrid, Spain}

\date{\today}
\pacs{75.25.Dk,61.72.-y,71.20.-b,91.60.Ed}

\begin{abstract}
We perform a detailed study of the electronic structure of Fe$_{1-x}$O at moderate values of $x$.
Our results evidence that the Fe vacancies introduce significant local modifications of the structural,
electronic and magnetic features, that serve to explain the origin of the measured dependencies of
the physical properties on $x$. 
The final properties are determined by a complex interplay of
the charge demand from O, the magnetic interactions, and the charge order at the Fe sublattice.
Furthermore, polaronic distributions of charge resembling those at magnetite, Fe$_3$O$_4$,
emerge for the most stable defect structures. This defines a unique scenario to understand the nature 
of the short-range correlations in Fe$_3$O$_4$, and unveils their intimate connection to the long-range 
charge order developed below the Verwey transition temperature.
\end{abstract}

\maketitle

\section{Introduction}

FeO belongs to the magnetic transition metal monoxides (TMO) series, together with MnO, CoO and NiO. 
They share similar properties, such as a cubic NaCl crystal lattice, large insulating gaps, or antiferromagnetism 
of AF-II type, where adjacent ferromagnetic (111) cation planes couple antiferromagnetically.
In spite of their apparent simplicity, the TMO exhibit intricate electronic interactions, and
have been considered as prototypes to explore electronic correlations in the verge
between Mott-Hubbard and charge transfer insulators \cite{Zaanen-1985}.
This has motivated a large number of calculations of their properties within {\it ab initio} frameworks 
\cite{Anisimov-1991,Dane-2009,Thunstrom-2012,Rodl-2009}.
Furthermore, all of them show a distortion of the cubic symmetry below the N\'eel temperature (T$_N$) 
\cite{Willis-1953,Roth-1958}, which has been unequivocally related to the magnetic exchange coupling 
constants \cite{Kant-2012}.
In the particular case of FeO, the distortion is rhombohedral, and consists in a slight elongation
along the $[111]$ direction \cite{Tombs-1950}.
%
%
The local magnetic moments align parallel to this direction
\cite{Shull-1951}, though a weak perpendicular component compatible with a departure from collinear spin order
has been solved \cite{Fjellvag-1996,Saines-2013}. 

Within the TMO series, FeO is singular in that it shows a large deficiency in Fe ranging from 5 to $15 \%$ 
(Fe$_{1-x}$O, with $x=0.05-0.15$) at ambient conditions. The stoichiometric form is stable at the extreme pressures and 
temperatures of the Earth's lower mantle, one of whose major constituents are solid solutions of the mineral 
form of FeO, w\"ustite. In fact, FeO is under scrutiny in the fields of geophysics and 
paleomagnetism, where large efforts are devoted to understand its complex
transitions at high pressures and to determine the distribution of Fe vacancies (V$_{Fe}$) in the lattice
\cite{Fei-1994,Ohta-2012,Hazen-1984}. 
Though unstable FeO decomposes in Fe$_3$O$_4$ and Fe below 570$^{\circ}$C, at room temperature 
the reaction is slow \cite{Willis-1953}, allowing to prepare Fe$_{1-x}$O by rapid quenching. The 
exact composition and local order of the defects is determined in a complex way both by the
preparation conditions and the history of the sample, and furthermore, may evolve with time \cite{Hazen-1984}.
However, it has been possible to identify
trends revealing the influence of V$_{Fe}$ on the physical properties of Fe$_{1-x}$O. 
For example, an empirical linear relation between the unit cell parameter $a$ and $x$ has been determined
\cite{McCammon-1984,Dimitrov-2000}. The magnetically induced rhombohedral distortion is also known to
decrease as the Fe content decreases \cite{Willis-1953,Fjellvag-1996}, and similarly the value of T$_N$ and 
the specific heat anomaly at the transition depend on $x$ \cite{Schrettle-2012}.
The modification of the heat capacity across T$_N$ evidences changes in the nature of the magnetic transition, 
with a much stronger cooperative character in FeO than in non-stoichiometric samples \cite{Fjellvag-1996}.
Also the conductivity and cation self-diffusion have been observed to depend on $x$, and the
carrier type changes from $p$ to $n$ for $x \sim 0.08$ \cite{Johnson-1969}.

These dependencies and the high values of $x$ admitted by w\"ustite suggest that Fe vacancies
arrange forming specific defect structures. Their identification has been a very active field of research 
since the pioneering neutron diffraction study of Roth \cite{Roth-1960}, and though it must be ultimately 
admitted that different local structures and long-range orders may exist depending on the history of a
sample \cite{Hazen-1984}, it seems clear that there is a relation between the V$_{Fe}$ and the emergence of
interstitial Fe$^{3+}$ cations \cite{Cheetham-1971,Press-1987}. 
Simple charge compensating arguments propose that creating an iron vacancy generates two octahedral Fe$^{3+}$,
as has been proved from first-principles calculations of isolated V$_{Fe}$ \cite{Wdowik-2013}.
However, both diffraction experiments and atomistic calculations agree that the minimal defect cluster is
formed by 4 V$_{Fe}$ surrounding a tetrahedral Fe$^{3+}$, the so-called 4:1 cluster \cite{Koch-1969,Catlow-1975}
(see figure \ref{111-distortion}).
The aggregation of these minimal units as $x$ increases may proceed in several ways, by face-, edge- or
corner-sharing of neighbouring 4:1 clusters. Though it is believed that the ratio between V$_{Fe}$ and
interstitials increases with temperature and at lower defect concentrations \cite{Press-1987}, different types
of aggregates may coexist \cite{Lebreton-1983}, and it is not even clear if spinel-type local structures
are favored over more compact aggregations \cite{Minervini-1999}.

Understanding the fundamental properties of defective Fe$_{1-x}$O at the atomic scale seems essential to 
determine the origin of the observed dependencies of the physical properties on $x$.
Surprisingly, however, detailed studies of the influence of V$_{Fe}$ 
on the electronic structure are scarce, and to the best of our knowledge, only isolated vacancies
have been considered with {\it ab initio} methods \cite{Wdowik-2013}. Here, we perform first-principles 
calculations of Fe$_{1-x}$O to determine the stability and electronic properties of 4:1 clusters, comparing
them to those of isolated V$_{Fe}$ and stoichiometric FeO. Furthermore, as the 4:1 cluster is a basic unit 
for the development of spinel structures at the FeO lattice, 
our study provides a bridge to explore the evolution from Fe$_{1-x}$O to Fe$_3$O$_4$, and the
influence of the lattice symmetry on the local order.
As we will show, this allows to extract interesting conclusions about the nature
of short-range charge correlations in Fe$_3$O$_4$.

The paper is organized as follows. After describing the methodology, we provide our results for the electronic
properties of FeO and isolated Fe vacancies in Fe$_{0.97}$O. Then, we describe unit cells with 3 V$_{Fe}$,
corresponding to a composition Fe$_{0.906}$O, comparing two different situations: disperse V$_{Fe}$ and
4:1 clusters. In the last section we explore the similitudes between the charge distributions 
of Fe$_{1-x}$O and Fe$_3$O$_4$ based on our previous calculations of magnetite \cite{Bernal-2014}.

\section{Computational method}

We have performed {\it ab initio} calculations within density functional theory (DFT) as implemented in the 
VASP code \cite{vasp1,vasp2}, using the projector-augmented wave (PAW) method for the treatment of the core 
electrons \cite{paw1,paw2} and the Perdew-Burke-Ernzerhof parametrization of the generalized gradient 
approximation modified for solids (PBEsol) for the exchange-correlation functional \cite{pbesol}. 
A kinetic energy cutoff of 400 eV has been used, and all symmetrizations have been removed. 
To take into account the important electronic correlations
in Fe oxides, we have included an on-site Coulomb repulsion term U following the Dudarev approach 
\cite{Dudarev-1998}.
Our choice of an effective $U-J=4$ eV is based on the recovery of the experimental values of both the 
equilibrium NaCl lattice parameter ($a= 4.30$ $\AA$) and the insulating gap ($E_g=2.0$ eV) for
stoichiometric FeO, while preserving the overlap and orbital order of the oxygen $p$ band and the 
cation $d$ states \cite{Fujimori-1987,Rodl-2009}. A similar procedure 
applied to Fe$_3$O$_4$ led us to the same choice of U in a previous study \cite{Bernal-topreview}, which eases 
comparison of the results corresponding to both materials.

\begin{figure}[thbp]
\begin{center}
\includegraphics[width=\columnwidth,clip]{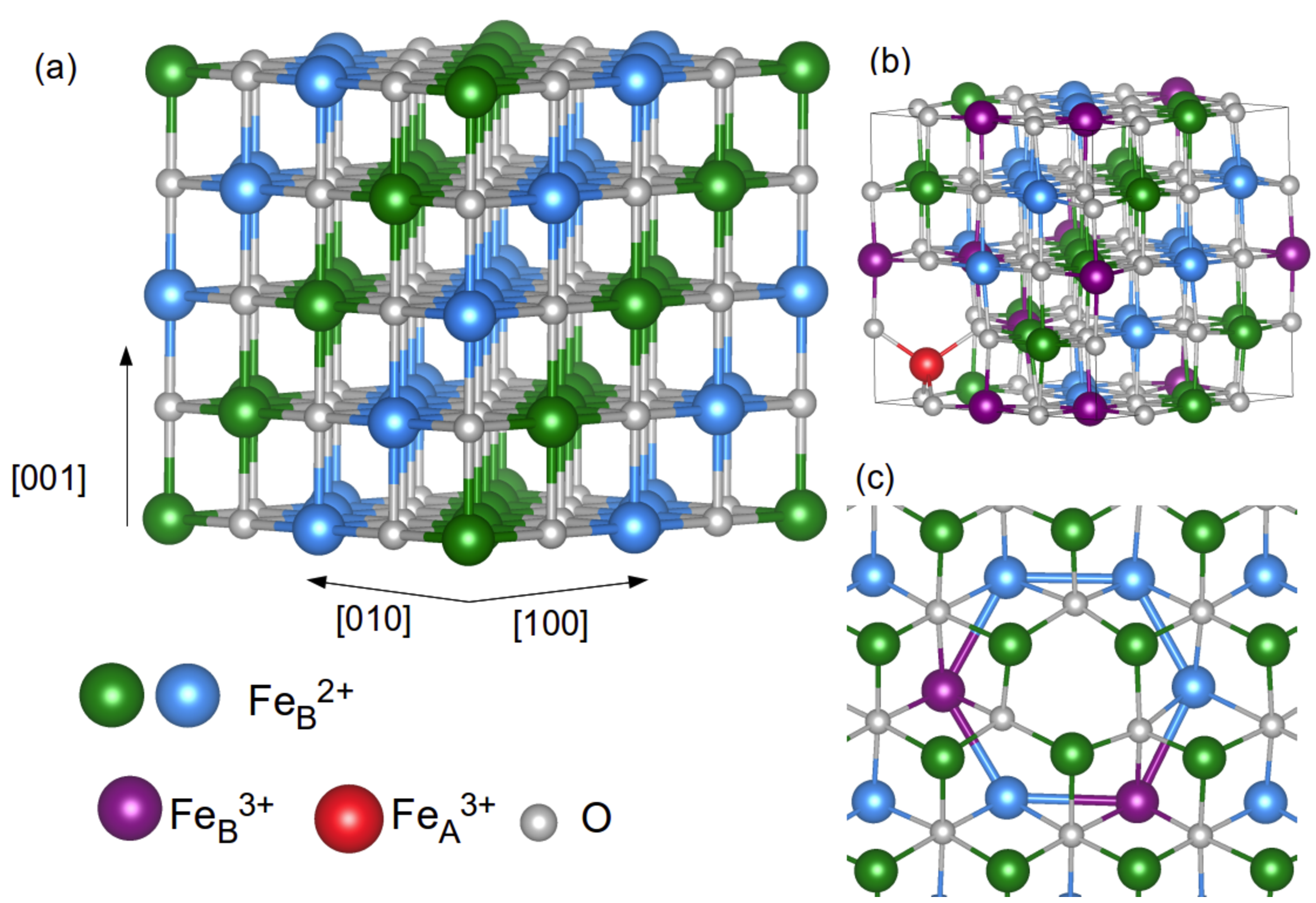}
\caption{(Color online).
(a) Cubic unit cell of 64 atoms used to model FeO. Fe sites with opposite spin orientation are coloured in blue and green.
(b) Same for a supercell with a 4:1 cluster. 
(c) Top view of three (111) planes around an Fe vacancy in Fe$_{0.97}$O. The central thick hexagon
joins the Fe atoms surrounding the vacancy site in the blue layer. 
\label{111-distortion}}
\end{center}
\end{figure}

Stoichiometric FeO can be modelled with a minimal unit cell of 4 atoms under AF-II order. Here we have used
a larger cubic unit cell of 64 atoms that allows to consider isolated 4:1 clusters as well as up to 3
separated V$_{Fe}$.
A sketch of the cubic unit cell is represented in the left panel of figure \ref{111-distortion}, where blue and green layers
denote Fe atoms with opposite spin orientation.
Full relaxation of the lattice vectors and atomic positions has been performed until the forces
on all atoms were lower than 0.01 eV/$\text{\AA}$. 
The Brillouin Zone (BZ) has been sampled using the Monhorst-Pack method, with partitions of 
$2 \times 2 \times 2$ during relaxations and $4 \times 4 \times 4$ for static calculations. This scheme 
guarantees a convergence in the total energy of 1 meV/atom.
When modelling FeO with the reduced cell of 4 atoms, a sampling of the BZ of $7 \times 7 \times 7$
has been used.

The energy cost of creating a defect has been estimated as \cite{VandeWalle-2004}:
\begin{equation}
E_c = E[Fe_{1-x}O] - E[FeO] + n \mu_{Fe},
\label{eq1}
\end{equation}
where $E$ refers to the total energy of the supercell with (Fe$_{1-x}$O) or without (FeO) defects,
$n$ to the number of V$_{Fe}$ introduced in the supercell, and $\mu_{Fe}$ to the chemical potential of Fe.
The limits of $\mu_{Fe}$ have been obtained from calculations of the total energy of bcc Fe and the O$_2$ 
molecule, taking into account the relation between the Gibbs free energies of formation of the oxide and 
the stable forms of its elemental constituents, and the influence of U \cite{Kresse-2014}.
This allows to express $E_c$ as a function of the O chemical potential $\mu_O$, where 
we have set the zero reference state to $\mu_O=1/2 E[O_2]$.

\section{Stoichiometric $\text{FeO}$}

As mentioned in the Introduction, the electronic structure of FeO has been widely studied previously
under diverse approaches, and the system has been often used as a test-bed for models aiming to
capture correlation effects \cite{Rodl-2009,Mazin-1997}.
Our description of stoichiometric FeO using the reduced unit cell of 4 atoms is in excellent agreement with 
previous calculations. 
When the large cube of 64 atoms is employed, the global features of the ground state do not vary,
leading to an identical value of the cubic lattice parameter $a=4.30$ $\AA$, similar 
electronic properties, and an AF-II order of the local Fe magnetic moments (3.7 $\mu_B$) that induces a comparable
rhombohedral distortion. However, under the reduced symmetry of this large unit cell a significant 
deformation of the O fcc sublattice manifests, enhancing the dispersion of the atomically resolved properties, 
such as local charges and interatomic distances.
The structural deformation already exists in the minimum unit cell of 4 atoms, though there it does
not alter the symmetrical distribution of the Bader charges.
The situation is reminiscent to that occuring in magnetite, 
as a hint of the  close relation between the properties of both Fe oxides. 
The distortion of the O sublattice lowers the total energy by 21 meV/f.u., and is related to the onset of magnetism: 
it does not emerge in a non-magnetic calculation, where also the elongation along $[111]$ is lost.

\begin{figure}[hbtp]
\begin{center}
\includegraphics[width=\columnwidth,clip]{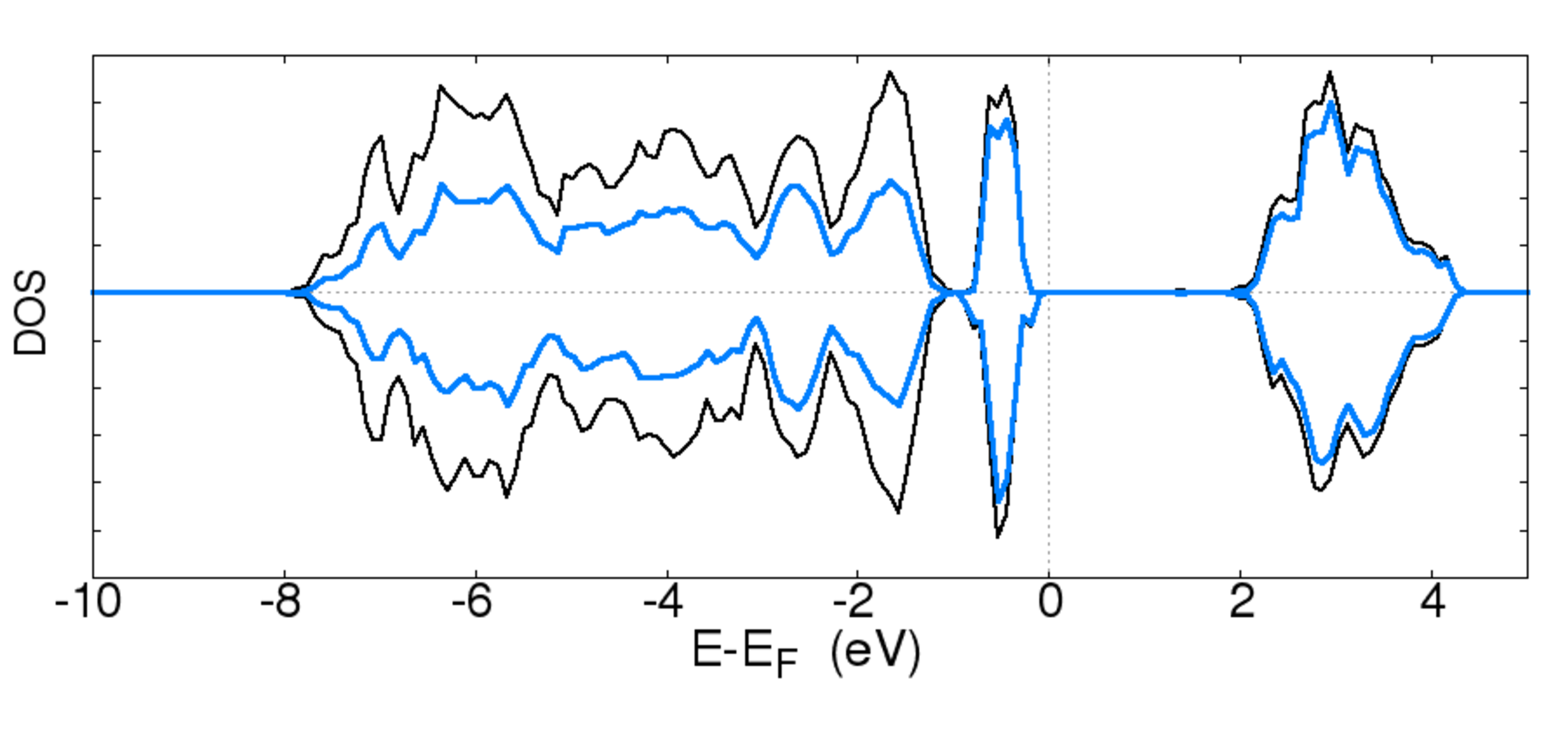}
\caption{(Color online).
Spin-resolved DOS of stoichiometric FeO with positive (negative) values corresponding to majority (minority) spins. 
The black curve corresponds to the total DOS of the supercell, the blue one to the Fe contribution.}
\label{FeO-full}
\end{center}
\end{figure}

\begin{figure}[thbp]
\begin{center}
\includegraphics[width=\columnwidth,clip]{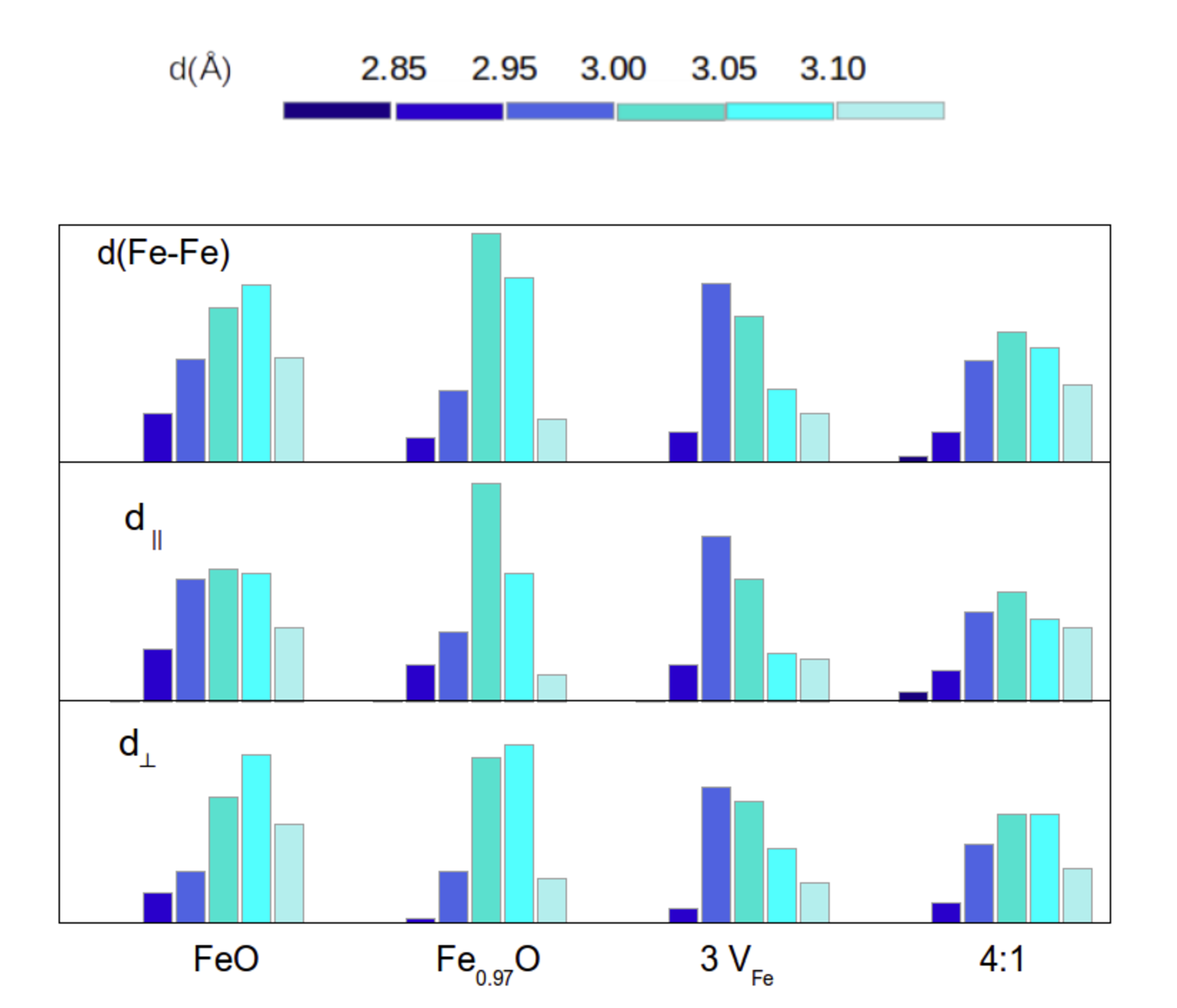}
\caption{(Color online).
Histograms showing the distribution of values of d(Fe-Fe) at the Fe$_B$ sublattice
and its decomposition in d$_\parallel$ and d$_\perp$ for the structures in table \protect\ref{charges}. 
The vertical scale of the middle and bottom panels has been reduced by $\sim 30\%$ with respect to
the top panel.
\label{histogram}}
\end{center}
\end{figure}

The total density of states (DOS) of the supercell and its projection on the Fe atoms are in figure \ref{FeO-full}.
Although our description of FeO corresponds to a Mott insulator, there is a strong hybridization between 
O and Fe states throughout the valence band (VB), with a non-negligible contribution of O at the VB edge.
The orbital projection on the Fe sites shows the expected separation between $t_{2g}$ and $e_g$
states linked to the local octahedral coordination. Minor variations in the DOS are found at different sites, 
as a consequence of the asymmetry in the O sublattice, but the dispersion of the local charges is moderate, 
as evidenced in table \ref{charges}. 
The first column of the table provides the mean values and the corresponding dispersions of the Bader charge (Q$_B$)
and magnetic moment of O and Fe.
Fe atoms are labelled as Fe$^{2+}_B$ by analogy to Fe$_3$O$_4$, as justified in the next section.
The ionic nature of 
the Fe-O bonds manifests in the large charge transfer from Fe to O. On the other hand, the perfect compensation 
of the Fe magnetic moments with opposite orientations, already reflected in the DOS, leads to the negligible 
magnetization at the O sites.
\begin{table}[ht]
\begin{center}
\caption{Mean values and dispersions of the Bader charges (Q$_B$) and magnetic moments ($m$) for all distinct 
atomic sites at the different structures under study (see text for details). \label{charges}}
\renewcommand{\arraystretch}{1.5} 
\renewcommand{\tabcolsep}{0.4pc} 
\begin{tabular}{ccccc}
\hline \hline
Q$_B$ & FeO & Fe$_{0.97}$O & 3 V$_{Fe}$ & 4:1 \\
\hline
Fe$^{2+}_B$ & $6.74 \pm 0.04$ & $6.74 \pm 0.04$ & $6.72 \pm 0.03$ & $6.71 \pm 0.03$ \\
Fe$^{3+}_B$ &                 & $6.43 \pm 0.02$ & $6.39 \pm 0.02$ & $6.43 \pm 0.03$ \\
Fe$^{3+}_A$ &                 &                 &                 & $6.37 \pm 0.00$ \\
O           & $7.26 \pm 0.03$ & $7.24 \pm 0.05$ & $7.22 \pm 0.04$ & $7.23 \pm 0.04$ \\
\hline  \hline
$m (\mu_B)$   & FeO & Fe$_{0.97}$O & 3 V$_{Fe}$ & 4:1 \\
\hline
Fe$^{2+}_B$ & $3.67 \pm 0.00$ & $3.67 \pm 0.02$ & $3.67 \pm 0.02$ & $3.68 \pm 0.03$ \\
Fe$^{3+}_B$ &                 & $4.14 \pm 0.00$ & $4.15 \pm 0.02$ & $4.14 \pm 0.03$ \\
Fe$^{3+}_A$ &                 &                 &                 & $4.12 \pm 0.00$ \\
O           & $0.01 \pm 0.01$ & $0.03 \pm 0.02$ & $0.05 \pm 0.04$ & $0.05 \pm 0.04$ \\
\hline \hline
\end{tabular}
\end{center}
\end{table}
\begin{table}[ht]
\begin{center}
\caption{Mean values and dispersions of the in-plane (d$_{\parallel}$)
and interlayer (d$_{\perp}$) interatomic Fe-Fe distances corresponding to (111) planes, for
the structures in table \protect\ref{charges}.  \label{Fdistances}}
\renewcommand{\arraystretch}{1.5} 
\renewcommand{\tabcolsep}{0.2pc} 
\begin{tabular}{ccccc}
\hline \hline
d(Fe-Fe), $\AA$ & FeO & Fe$_{0.97}$O & 3 V$_{Fe}$ & 4:1 \\
\hline
d$_\parallel$ & $3.03 \pm 0.03$ & $3.03 \pm 0.05$ & $3.01 \pm 0.08$ & $3.03 \pm 0.08$ \\
d$_\perp$     & $3.05 \pm 0.01$ & $3.05 \pm 0.04$ & $3.02 \pm 0.06$ & $3.04 \pm 0.06$ \\
\hline  \hline
\end{tabular}
\end{center}
\end{table}
The structural properties are summarized in tables \ref{Fdistances} and \ref{Odistances}, and figure \ref{histogram}.
The rhombohedral distortion induced by the AF order stretches the cubic
cell along the $[111]$ axis perpendicular to the ferromagnetic (FM) Fe sheets, that we define as the $c$
direction. A measure of this distortion is
provided by the ratio of the mean interatomic distances between Fe neighbors at the same FM layer
(d$_\parallel$) and at adjacent layers (d$_\perp$), shown in the first column of table \ref{Fdistances}.
It is also evidenced from the more detailed distribution of values of d$_\parallel$ and d$_\perp$
in figure \ref{histogram}, peaked to lower values for d$_\parallel$ than for d$_\perp$.
A similar situation arises at the O sublattice, with identical mean values of d$_\parallel$ and d$_\perp$,
but larger dispersions reflecting the additional deformation of the sublattice.
In turn, this causes significant variations of the Fe-O bond lengths, d(Fe-O),
as reflected in the first column of table \ref{Odistances}.
However, the d(Fe-O) do not seem to follow any pattern with respect to the lattice or the magnetic order.
The noticeable distortion of the FeO lattice is closely linked to the competition between the different
superexchange paths. The magnetic order of the system is known to be dominated by the exchange
interactions between second nearest neighbors ($J_{2nn}$), due to frustration of the first nearest
neighbors interactions ($J_{1nn}$) \cite{Roth-1958}. Oppositely, the rhombohedral distortion of the unit
cell is caused by spin-phonon couplings governed by the $J_{1nn}$ \cite{Kant-2012}. The additional deformation 
of the O sublattice seems to be a mechanism to preserve maximization of the $J_{2nn}$ under the presence
of the magnetically induced phonon splitting.
\begin{table}[ht]
\begin{center}
\caption{Mean values and dispersions of the Fe-O bond lengths for the distinct types of Fe
sites at the structures in table \protect\ref{charges}.
\label{Odistances}}
\renewcommand{\arraystretch}{1.5} 
\renewcommand{\tabcolsep}{0.2pc} 
\begin{tabular}{ccccc}
\hline \hline
d(Fe-O), $\AA$ & FeO & Fe$_{0.97}$O & 3 V$_{Fe}$ & 4:1 \\
\hline
Fe$^{2+}_B$ & $2.15 \pm 0.05$ & $2.15 \pm 0.07$ & $2.14 \pm 0.06$ & $2.14 \pm 0.07$ \\
Fe$^{3+}_B$ &                 & $2.07 \pm 0.05$ & $2.05 \pm 0.07$ & $2.05 \pm 0.06$ \\
Fe$^{3+}_A$ &                 &                 &                 & $1.90 \pm 0.02$ \\
\hline  \hline
\end{tabular}
\end{center}
\end{table}

\section{Isolated iron vacancy: $\text{Fe$_{0.97}$O}$}

\begin{figure}[thbp]
\begin{center}
\includegraphics[width=\columnwidth,clip]{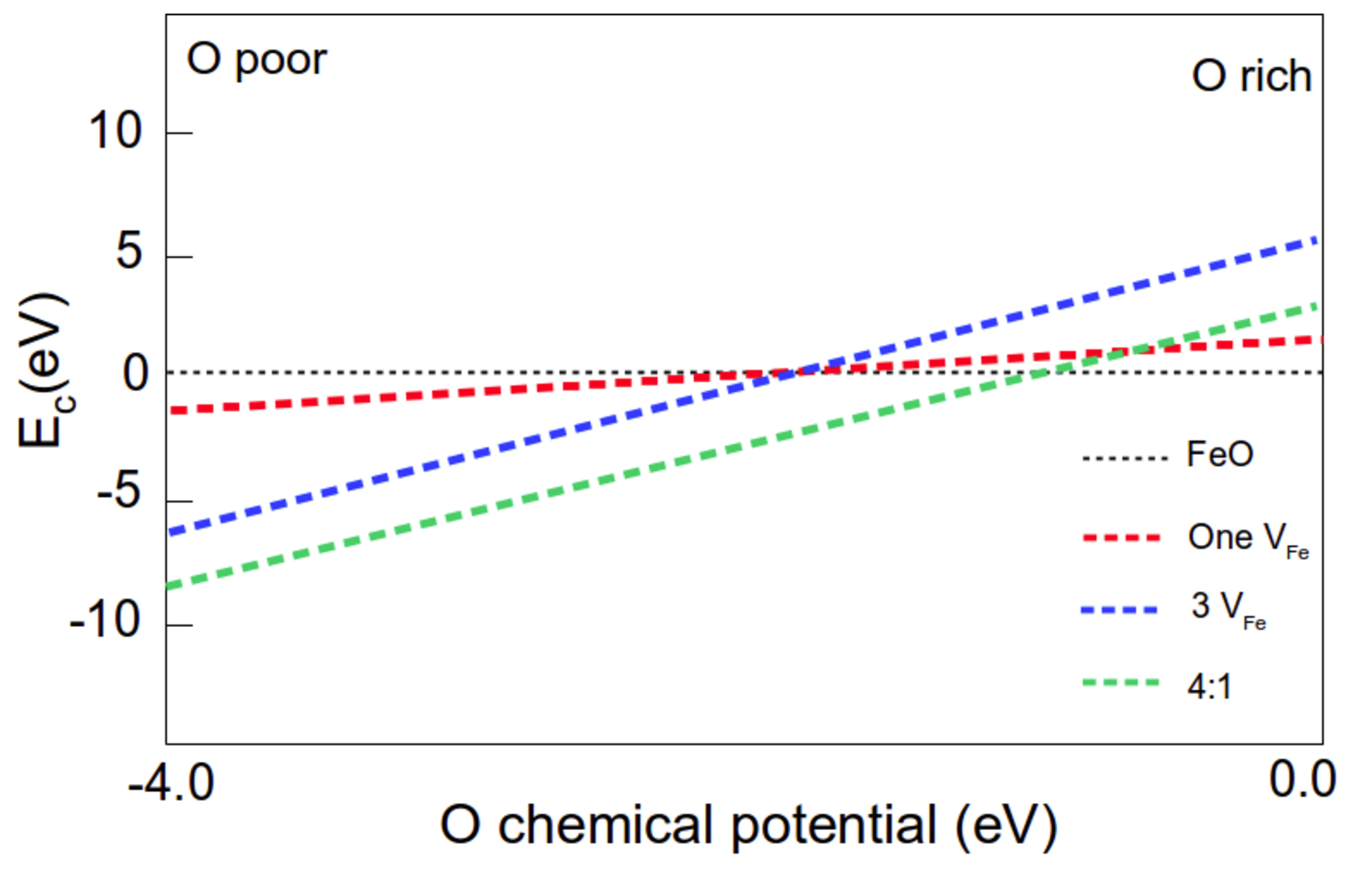}
\caption{(Color online).
Energy cost of creating a defect at the different structures considered in table \protect\ref{charges},
as a function of $\mu_O$.}
\label{fenergy}
\end{center}
\end{figure}

\begin{figure}[thbp]
\begin{center}
\includegraphics[width=\columnwidth,clip]{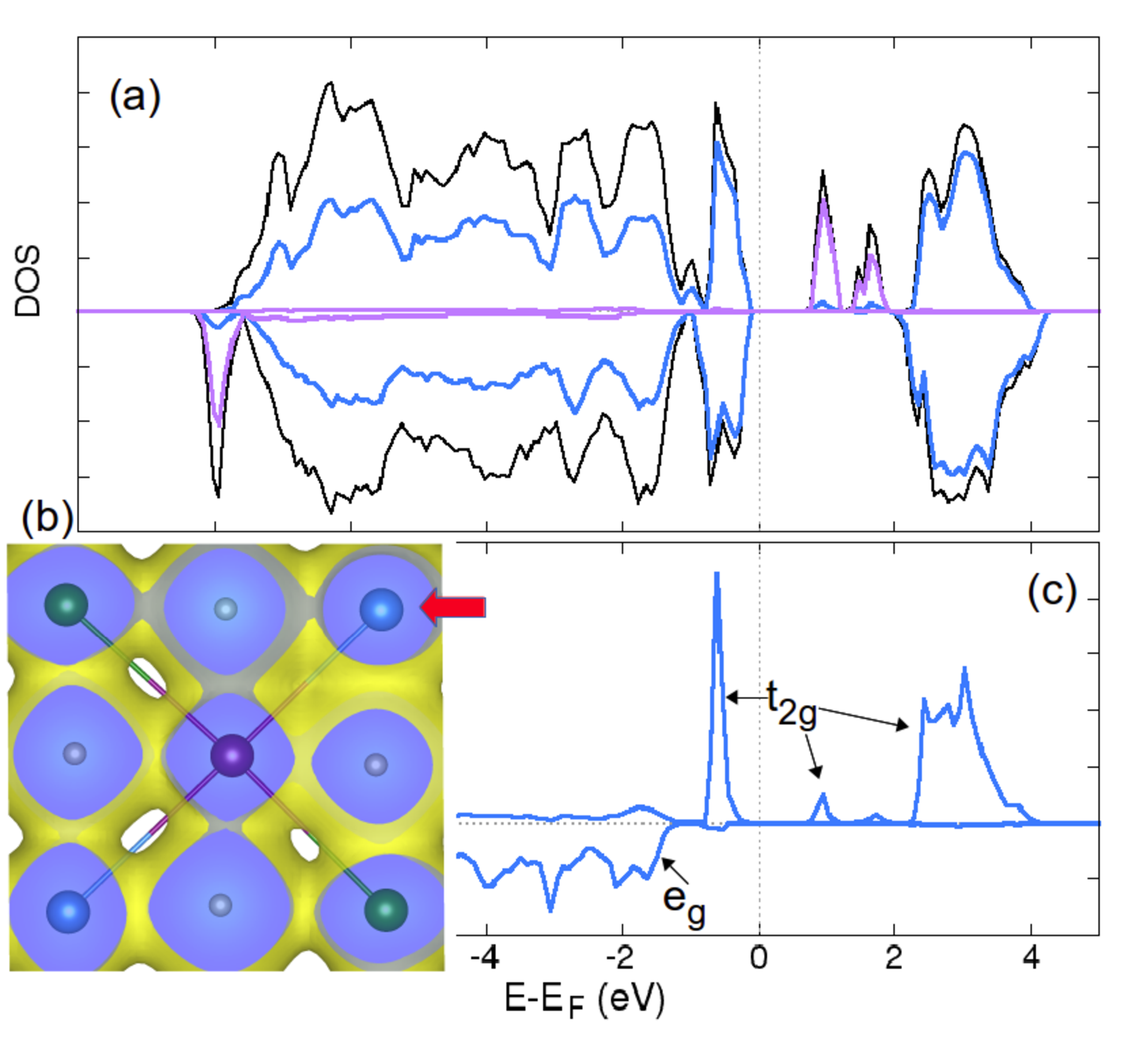}
\caption{(Color online).
(a) Total DOS of Fe$_{0.97}$O (black curve) with positive (negative) values corresponding to majority (minority) spins. The blue 
and purple curves are the projections on the Fe$^{2+}_B$ and Fe$^{3+}_B$ sites, respectively. (b) Charge density around an Fe$^{3+}_B$ site,
showing the accumulation of charge at certain Fe$^{3+}_B$-Fe$^{2+}_B$ connecting lines. (c) DOS around the Fermi level of the Fe$^{2+}_B$ pointed 
by an arrow in panel (b), showing the states directed along the CL that overlap with those of Fe$^{3+}_B$. 
The vertical scale is 10 times that in panel (a).}
\label{DOS-1vac}
\end{center}
\end{figure}

We have first considered the effect of creating one Fe vacancy in the cube of 64 atoms, 
which corresponds to $x=0.03$. This situation is outside the experimental range of concentrations 
found for w\"ustite, as we reproduce from our estimations of $E_c$ shown in figure \ref{fenergy}.
Its interest rather relies on the ability to isolate the features introduced by non-interacting vacancies.
In figure \ref{DOS-1vac} we show the total DOS of the supercell, together with the
projections on the two different types of Fe that can be distinguished. The corresponding mean
values and dispersions of the Bader charges and magnetic moments are in the second column of table \ref{charges}.
The emergence of Fe atoms that share more charge with O is evident. 
As mentioned in the introduction, creation of an Fe vacancy is expected to induce a change of valence on two Fe cations,
that should act as Fe$^{3+}$. We label them Fe$^{3+}_B$ by analogy to magnetite: in Fe$_3$O$_4$,
Fe cations order in octahedral (B) and tetrahedral (A) sublattices,
and at low temperatures a charge disproportionation at sublattice B leads to Fe atoms
acting either with valence $3+$ (Q$_B$=6.37) or $2+$ (Q$_B$=6.64) \cite{Bernal-2014,Verwey}. 
Here, the Fe$^{3+}_B$ sites are among the first neighbors of the Fe vacancy, and belong to the same FM layer
(see figure \ref{111-distortion}c). 
They introduce an incomplete charge compensation, leading to the slight reduction of the mean Q$_B$
of O with respect to FeO. In general, the 6 O atoms which are nearest neighbours (n.n.) to the vacancy show 
a lower charge than those farther from it, but the charge distribution is complex, and some O atoms far 
from the vacancy site hold similar low values of Q$_B$.

As evidenced in the table, accompanying the difference in Q$_B$
there is an important enhancement of the magnetic moment at the Fe$^{3+}_B$ sites, 
that also causes a slight increase of the induced O magnetization.
This leads to the partial compensation of the AF coupling at the supercell, unbalanced by the V$_{Fe}$.
However, a net magnetic moment of 2 $\mu_B$ yet remains, that together with the important 
reduction of the insulating gap to around 0.8 eV represent the main differences with respect to stoichiometric FeO. 
The new states at the gap are precisely those coming from the Fe$^{3+}_B$ states, that provide also the additional features 
at the bottom of the conduction band (CB).
But the global electronic structure of the FeO lattice does not vary significantly, and most modifications are localized 
around the vacancy site.

The presence of the V$_{Fe}$ introduces also changes in the structure. 
The second column of table \ref{Fdistances} summarizes the mean values of the interatomic distances between Fe neighbors,
d(Fe-Fe), that together with the distribution of their values in figure \ref{histogram} 
provide a measure of the rhombohedral distortion. 
As compared to stoichiometric FeO, there is a slight reduction of the cell volume 
and a tendency to shorten both d$_{\parallel}$ and d$_{\perp}$, better evidenced in figure \ref{histogram}. But
the elongation along $c$ is preserved for the low concentration of defects considered here. 
The relevant structural changes occur in the neighborhood of the vacancy. On one hand, they manifest
in the d(Fe-O), compiled in the second column of table \ref{Odistances}, which vary depending on the valence of Fe.
The O atoms tend to approach the Fe$^{3+}_B$ sites, while the average distance to Fe$^{2+}_B$ remains similar to 
the stoichiometric case. The noticeable dispersion of the values 
reflects that a significant distortion of the fcc O sublattice as obtained for FeO still exists.
On the other hand, there is an important rearrangement of the Fe positions in the vicinity of the V$_{Fe}$.
Due to the fcc symmetry of the Fe sublattice, each vacancy has 6 Fe n.n. at the (111) plane, 3 n.n. above and 3 n.n. below. 
There is a tendency to fill the void created by the vacancy, which causes a slight shrinkage of the 
hexagon defined by its in-plane Fe neighbors (see figure \ref{111-distortion}c) and also of the triangles defined
by its upper and lower n.n., which furthermore reduce their interlayer distances to the plane of the vacancy. 
We find that the d(Fe-Fe) of stoichiometric FeO are gradually restored from the second n.n. to V$_{Fe}$.
This was not observed in previous calculations \cite{Wdowik-2013}, probably linked to an incomplete
relaxation of the cubic symmetry.

Though the d(Fe-Fe) do not show the clear dependency with the Fe valence observed for d(Fe-O),
some of the Fe$^{2+}_B$ exhibit a singular relation to the neighboring Fe$^{3+}_B$, as indicated in
the lower panels of figure \ref{DOS-1vac}. Panel (b) represents the three dimensional real space distribution
of the charge density in the (010) plane defined by an Fe$^{3+}_B$ (central atom) and some of its Fe$^{2+}_B$ neighbors
(corner atoms). It is evident from the figure that there is a charge accumulation at one of the 
Fe$^{3+}_B$-Fe$^{2+}_B$ connecting lines (CLs). The d(Fe-Fe) defined by this CL is
considerably shortened to 2.88 $\AA$, to be compared with values over 3 $\AA$ for the rest of 
Fe$^{2+}_B$-Fe$^{3+}_B$ pairs in the figure. 
The charge accumulation at this short CL is also reflected in the DOS at figure \ref{DOS-1vac} (c),
that shows the states of the corresponding Fe$^{2+}_B$ overlapping those of Fe$^{3+}_B$ at the FeO gap. 
Analysis of the orbital character of these states indicates that they are
formed by t$_{2g}$ orbitals lying along the CL.
The combined presence of all these features is a signature of the short-range charge correlation units in magnetite
\cite{Bernal-2014}, the trimerons, formed by groups of three neighboring Fe$^{3+}_B$-Fe$^{2+}_B$-Fe$^{3+}_B$ sites with 
shortened interatomic distances and polaronic charge sharing \cite{Attfield-nature,Piekarz-2014}.
We will discuss these aspects in more detail in the last section.

\section{$\text{Fe$_{0.906}$O}$: Isolated vacancies}

\begin{figure}[thbp]
\begin{center}
\includegraphics[width=\columnwidth,clip]{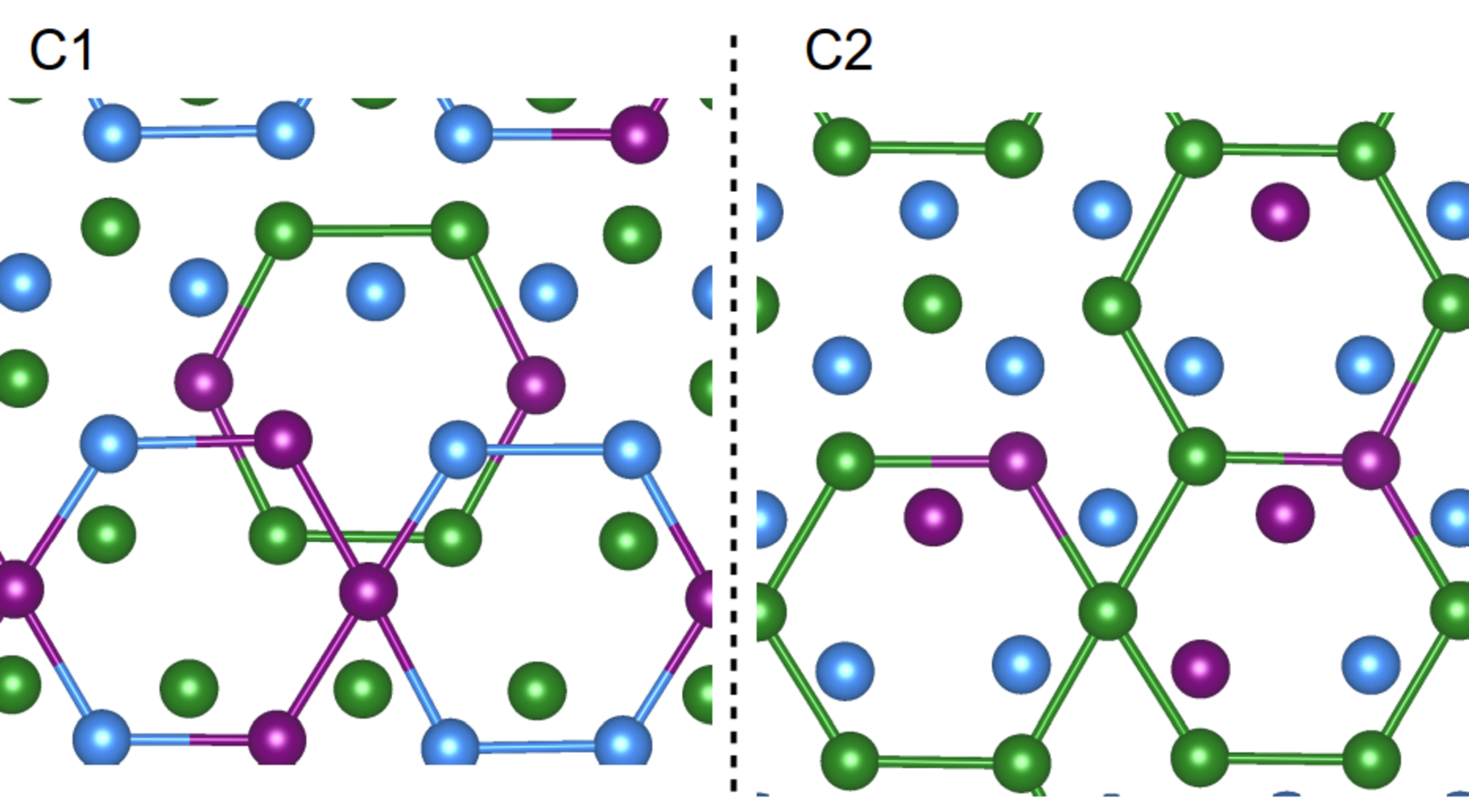}
\caption{(Color online).
Top view of two adjacent Fe(111) planes of Fe$_{0.906}$O with 3 isolated V$_{Fe}$, under configurations
C1 (left) and C2 (right). The lines join the six in-plane n.n. of each V$_{Fe}$.
The legend of figure \protect\ref{111-distortion} has been used, and the O sublattice has been omitted for clarity.
} 
\label{3vaclayers}
\end{center}
\end{figure}

When 3 V$_{Fe}$ are placed in the cube of 64 atoms, the maximum separation between them (around 6 $\AA$) 
can be achieved under two magnetically inequivalent configurations.
The situation is better visualized grouping all inequivalent Fe positions of the cube in two adjacent (111) 
planes of opposite spin orientation, as shown in figure \ref{3vaclayers}.
On the left side (case C1), two V$_{Fe}$ lie on the same plane (blue) and the other 
on the adjacent one (green), resulting in one uncompensated Fe magnetic moment.
On the right (case C2), all the vacancies lie on the same (111) plane, thus corresponding to removal of 3 Fe atoms with
parallel spins.
Surprisingly, 
in spite of the large difference in their net magnetizations (2 $\mu_B$ for C1, 14 $\mu_B$ for C2),
the total energies of both configurations are very close, the case C1 being favored by only 47 meV
(less than 1 meV/atom).
This tendency manifests also in the similarity of their global structural and electronic properties,
and eventhough we will show that there are local differences between both cases, their resemblance is a hint
of the non-interacting nature of the 3 V$_{Fe}$. This serves to put limits
on the distance for Fe vacancies to interact.
On the other hand, regarding figure \ref{fenergy}, this composition is more stable than FeO at low O$_2$ 
pressures, and also more stable than Fe$_{0.97}$O at the conditions where FeO is unstable,
evidencing a tendency to cluster the defects.

\begin{figure}[thbp]
\begin{center}
\includegraphics[width=\columnwidth,clip]{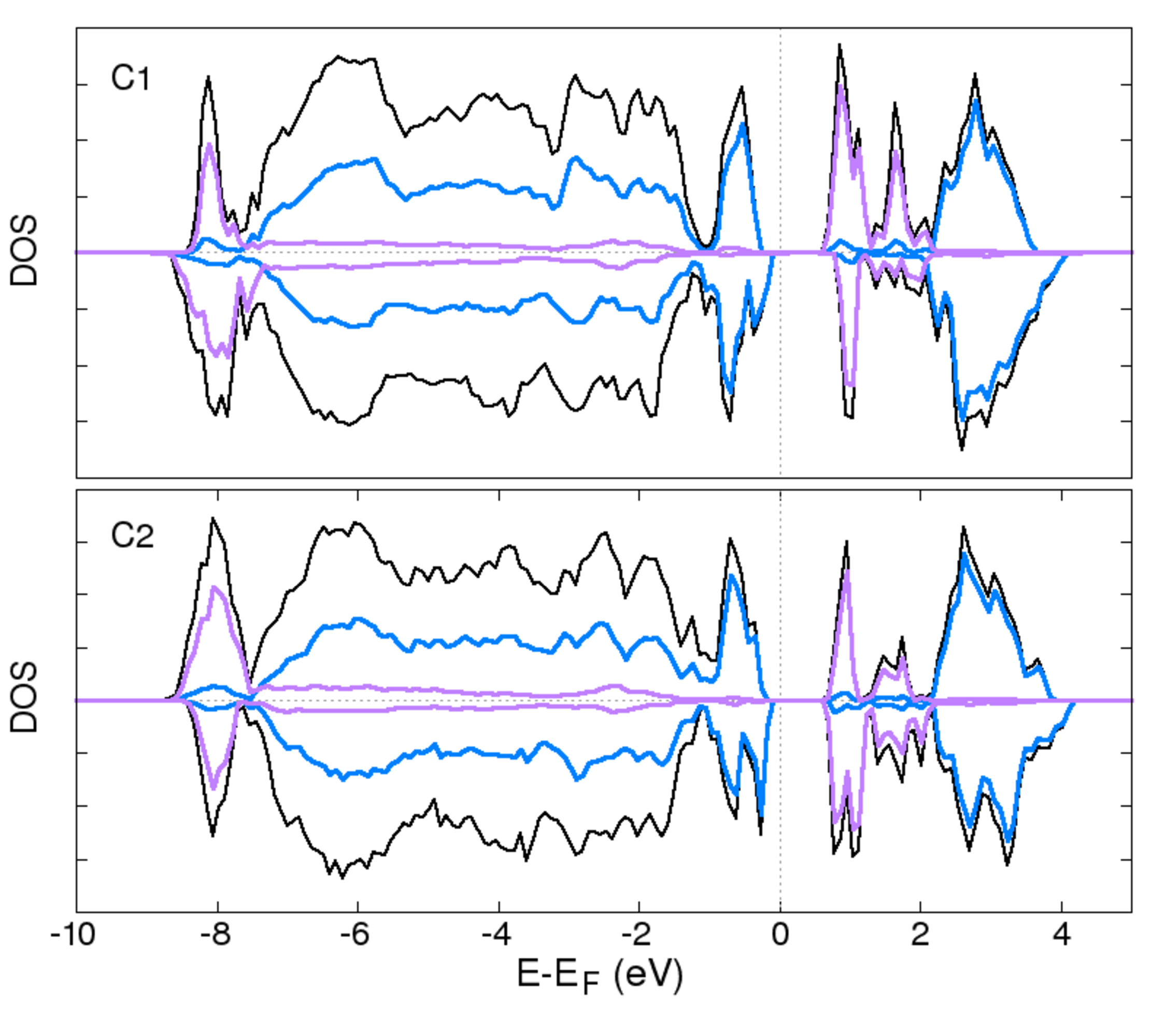}
\caption{(Color online).
Same as figure \protect\ref{DOS-1vac}a for the supercells modelling Fe$_{0.906}$O with 3 isolated V$_{Fe}$.
Top (bottom) panel corresponds to the C1 (C2) configuration.}
\label{DOS-3vac}
\end{center}
\end{figure}
The DOS of the supercells corresponding to C1 and C2 is shown in figure \ref{DOS-3vac}, together with the 
projections on the different types of cations. 
As occured for one isolated vacancy, in both configurations two Fe per vacancy
emerge with enhanced $3+$ valence. Again they introduce new states at the bottom CB and at the insulating gap,
reducing it to $\sim 0.5$ eV. 
Comparing to the DOS of Fe$_{0.97}$O in figure \ref{DOS-1vac}, the larger number of Fe$^{3+}_B$ sites introduces 
additional states, that moreover appear at both spin orientations. In general all features of the DOS are broader,
associated to the more defective structures. 

The mean values of the charges and magnetic moments are summarized in the third column of table
\ref{charges}. They correspond to C1, but similar values are obtained for C2.
As compared to Fe$_{0.97}$O, the increased number of V$_{Fe}$ enhances the demand of charge from O, resulting
in reduced Bader charges of both the Fe and O atoms.
Also the Fe$^{3+}_B$ are not only first neighbors to one V$_{Fe}$, but sometimes to two of them,
which reflects in their reduced Q$_B$. 
As the Fe$^{+3}_B$ atoms emerge to compensate the loss of charge donors introduced by the vacancies,
not all of them lie on the same (111) plane of the closest V$_{Fe}$. 
Their enhanced magnetic moments thus reinforce the spin imbalance, an effect more pronounced in C2:
while in C1 there is an equal number of Fe$^{3+}_B$ sites with each spin orientation,
in C2 two Fe$^{3+}_B$ are in the plane of the V$_{Fe}$, and four in the adjacent one.
This combined to the small relative energy difference between C1 and C2 supports the interpretation of
magnetic measurements, that assign to the defects local ferrimagnetic areas which are disordered with respect to 
each other and also with respect to the overall AF order \cite{Fjellvag-1996,Battle-1979}.

In the third column of tables \ref{Fdistances} and \ref{Odistances} we show the mean values of the 
d(Fe-Fe) and d(Fe-O) for configuration C1.
Again these global values, together with the unit cell vectors and cell volume, are very similar in C1 and C2.
The volume contraction with respect to FeO is 3 times that found for Fe$_{0.97}$O,
corresponding to a cubic lattice parameter of 4.26 $\AA$. 
The average values of d(Fe-O) are also slightly shorter
than for Fe$_{0.97}$O, reflecting the large demand of charge from O due to the increased number of V$_{Fe}$.
Furthermore, there is
a weakening of the rhombohedral distortion, in agreement with the experimental evidence
\cite{Willis-1953,Fjellvag-1996}.
This is better seen in the distribution of values of d(Fe-Fe) in figure \ref{histogram}, very similar
for d$_{\parallel}$ and d$_{\perp}$. The large number of vacancies and the tendency of the Fe atoms to fill
their voids shifts all distances to shorter values.
Another important effect is that some pairs of Fe$^{3+}_B$ are now first neighbors, inducing an accumulation
of the valence charge in the interstitial region between them.  
This tends to inhibit the polaronic charge sharing between Fe$^{3+}_B$-Fe$^{2+}_B$ pairs of shortened d(Fe-Fe)
observed in the dilute limit $x=0.03$. Regarding figure \ref{DOS-3vac}, here there are still Fe$^{2+}_B$ states 
that overlap those of Fe$^{3+}_B$ at the FeO gap. They are always linked to values of d(Fe-Fe) 
below 2.97 $\AA$ and to the existence of orbital order along the corresponding CL. However,
no charge accumulation occurs at the interstitital region between Fe$^{2+}_B$ and Fe$^{3+}_B$.
The additional requirement for polaronic charge sharing is the shortening of the
Fe$^{2+}_B$-Fe$^{3+}_B$ interatomic distance below 2.89 $\AA$, that is strongly suppressed
by the accumulation of close Fe$^{3+}_B$ sites.

In summary, the high concentration of V$_{Fe}$ in Fe$_{0.906}$O starts to break the homogeneity 
of the FeO matrix. Though each vacancy only modifies the local area surrounding it, and 
the interaction between vacancies is weak,
the structure of FeO cannot accommodate such a large number of defects
without altering its properties. This has important consequences for the charge distribution.

\section{$\text{Fe$_{0.906}$O}$: 4:1 clusters}

The composition Fe$_{0.906}$O can also be achieved by grouping the V$_{Fe}$ in defect clusters.
The most compact one is the 4:1 cluster, where we introduce an interstitial Fe at a 
tetrahedral coordination site, Fe$_A$, and surround it by 4 V$_{Fe}$. 
As one of the vacancy sites at the octahedral Fe sublattice is balanced by the creation of Fe$_A$, 
the stoichiometry corresponds to the effective introduction of 3 V$_{Fe}$, similarly to the case studied above.
The 4:1 cluster is expected to be the most stable configuration for moderate
values of $x$ \cite{Minervini-1999}, and actually we obtain that it lowers the energy by 22 meV/atom as compared
to the supercell with 3 isolated V$_{Fe}$. 
In turn, this extends the range of stability of defective Fe$_{1-x}$O over FeO to higher O pressures, as
shown in figure \ref{fenergy}.

\begin{figure}[thbp]
\begin{center}
\includegraphics[width=\columnwidth,clip]{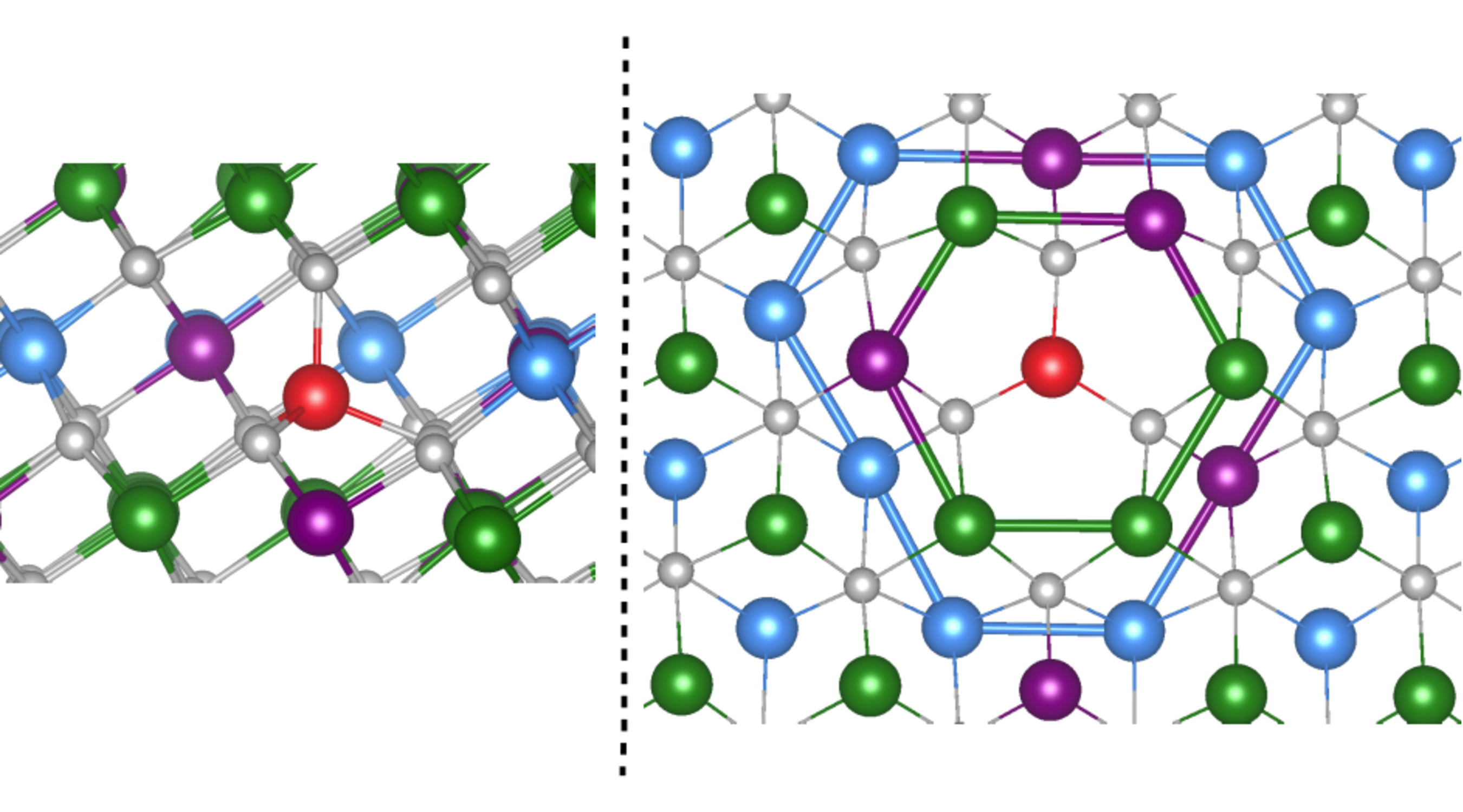}
\caption{(Color online).
Side view along $c$ (left) and top view (right) of Fe$_{0.906}$O around a
4:1 cluster, following the legend in figure \protect\ref{111-distortion}. 
On the right, only two adjacent (111) Fe layers are shown, and
the Fe$_A$ atom lies between them. The thick lines join the in-plane n.n. to the vacancy site.}
\label{layer4x1}
\end{center}
\end{figure}

Figure \ref{layer4x1} represents the 4:1 cluster embedded in our supercell, together with a top view of the (111) 
Fe layers around Fe$_A$, identifying the vacancy sites.
It is evident from the figure that three out of the four V$_{Fe}$ are first n.n. lying on the same (111) plane.
Since the spin of Fe$_A$ can be parallel or antiparallel to the layer with more vacancies, two magnetic configurations
can be envisaged, with a very different uncompensated magnetization.
As occured for 3 isolated V$_{Fe}$, the corresponding ground states show similar global 
properties except for a very different net magnetic moment, of either 4 $\mu_B$ or 14 $\mu_B$.
The relative energy difference between both configurations is considerably higher than between C1 and C2, 
but still it is only moderate, of 5 meV/atom, and favors the situation with the lower magnetization.
In fact most differences are similar to those discussed for C1 and C2, so in the following 
we will just describe the features of the most stable case.

\begin{figure}[thbp]
\begin{center}
\includegraphics[width=\columnwidth,clip]{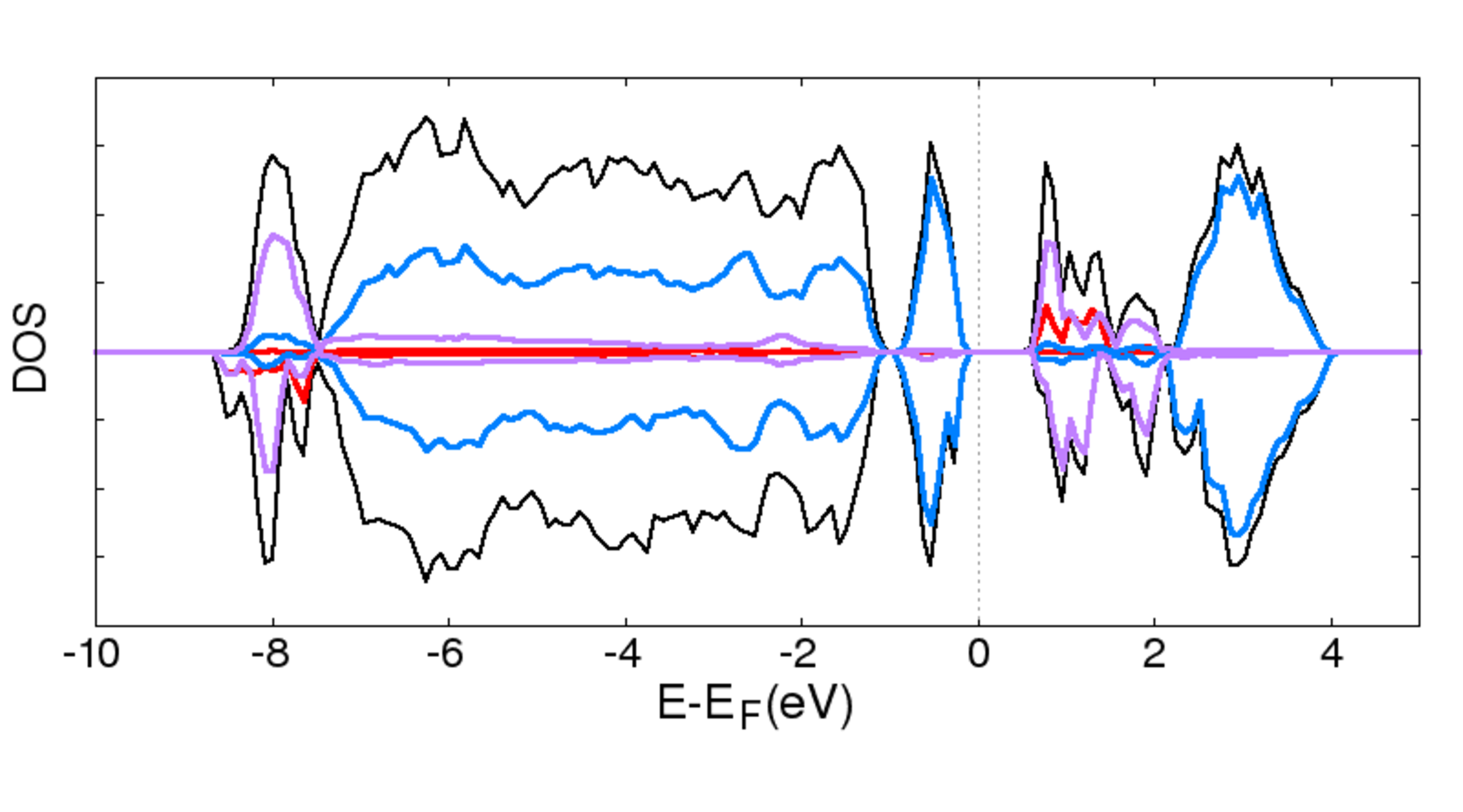}
\caption{(Color online).
Same as figure \protect\ref{DOS-1vac}a for the supercell modelling 4:1 clusters in Fe$_{0.906}$O.
The red curve is the projection on the tetrahedral Fe$^{3+}_A$ atom.}
\label{4x1DOS}
\end{center}
\end{figure}

The resulting DOS of the supercell and the projections on the different Fe sites
are shown in figure \ref{4x1DOS}, while the mean values of the Q$_B$ and magnetic moments 
are summarized in the last column of Table \ref{charges}. 
The charge defect introduced by the 3 effective V$_{Fe}$ would require 6 octahedral Fe$^{3+}_B$ atoms.
However, as Fe$_A$ also acts with $3+$ valence, only 5 Fe$^{3+}_B$ atoms emerge. All of them are
first n.n. to two V$_{Fe}$ simultaneously, and they are also among the closest octahedral Fe neighbors of Fe$_A$,
as shown in figure \ref{layer4x1}.
The tetrahedral atom introduces additional states at the bottom conduction and valence bands, that overlap 
the contribution from Fe$^{3+}_B$, and the resulting gap is $\sim 0.6$ eV. 
Compared to cases C1 or C2, the distinct features corresponding to each type of Fe site in the DOS
are better resolved, which evidences a higher degree of order in the structure.
For example, the peaks associated to the $t_{2g}$ and $e_{g}$ states of Fe$^{2+}_B$ at the top VB do not overlap,
similarly to stoichiometric FeO, and opposite to all previous defect structures considered above.
The higher localization of the defects has also consequences in the distribution of the O charges,
which are lower (around 7.18) for the atoms bonded to Fe$_A$, and higher far from the defect,
where they reach values similar to stoichiometric FeO. 
The result is that the mean Q$_B$ at the O sublattice is enhanced over the configuration corresponding to 
3 isolated V$_{Fe}$. This effectiveness of the 4:1 cluster to globally compensate the loss of charge introduced by the V$_{Fe}$
seems to be at the origin of its higher stability.

The magnetic moments of Fe$^{3+}_B$ and Fe$^{3+}_A$ show less differences than their Q$_B$, and the 
induced magnetization at the O sublattice is similar for all distributions of V$_{Fe}$ in Fe$_{0.906}$O.
On the other hand, the formation of the clusters leads to the inhomogeneous distribution of the magnetic moments,
with a higher magnetization in the region around the defects. This is in excellent agreement with previous 
experiments \cite{Dimitrov-1999}, where high magnetization and low temperature coercivity have been
linked to the presence of large clusters based on stacking spinel-like defects.

The mean values and dispersions of d(Fe-Fe) and d(Fe-O) are summarized in the last columns of tables \ref{Fdistances} 
and \ref{Odistances}. The d(Fe-O) for the octahedral Fe sites do not differ substantially from the case 
corresponding to 3 isolated V$_{Fe}$, but there is an important shortening of the bond lengths 
corresponding to Fe$_A$.
The values are close to those found at the tetrahedral sublattice in magnetite \cite{Bernal-2014}.
Oppositely, important differences can be found in the d(Fe-Fe) depending on the arrangement of the 
V$_{Fe}$. As evidenced in figure \ref{histogram}, the range of interatomic distances is wider 
in the presence of the 4:1 cluster,
and their distribution closer to stoichiometric FeO than for any other defect structure in the figure. 
The largest differences to the stoichiometric case come from the increased similitude between d$_{\parallel}$ 
and d$_{\perp}$, and from the global shortening of the d(Fe-Fe), including the emergence of particularly 
low values below 2.85 $\AA$. 
On one hand, this reflects the volume contraction to fill the voids left by the vacancies.
On the other, the decrease of the magnetically induced rhombohedral distortion upon decrease of the Fe content,
in good agreement with the experimental evidence.
But in general, the FeO matrix tends to recover its structure far from the V$_{Fe}$,
while the defect cluster introduces larger local modifications than the isolated vacancies.
Furthermore, the shortest d(Fe-Fe) are signatures of the emergence of local polaronic charge distributions,
as will be demonstrated in the next section. 

\section{Charge density distribution. Comparison to $\text{Fe$_3$O$_4$}$.}

Magnetite, Fe$_3$O$_4$, is one of the stable forms of binary Fe oxides. It is a material of interest for magnetic
applications, due to its unique properties of large magnetization, high magnetic ordering temperature
and half-metallicity. 
Furthermore, at T$_V$=120 K it undergoes a largely studied metal-insulator transition, the Verwey transition 
\cite{Verwey}, where the interplay between structure and charge-order (CO) is still under debate 
\cite{Walz-2002,Garcia-2004}.
At ambient conditions the crystal structure is the cubic inverse spinel, 
that below T$_V$ transforms into a large monoclinic unit cell accompanied by a drop of the conductivity and
the emergence of CO at the octahedral Fe sublattice.
Understanding the features of the CO in Fe$_3$O$_4$ has proved to be a complex task 
\cite{Leonov-2004,Piekarz-2007,Garcia-2009,Weng-2012},
and the survivance of short-range correlations above T$_V$ linked to the
long-range CO has been recently demonstrated \cite{Piekarz-2014}.
The local order generated by the 4:1 clusters in Fe$_{1-x}$O is a basic building block for the development of a spinel-like
structure similar to that of magnetite. Moreover, the local properties around the defect are very similar
to those of Fe$_3$O$_4$. Here we will explore how these similarities manifest also in the charge distribution,
converting Fe$_{1-x}$O in a unique material to understand the features of the CO of magnetite.

\begin{figure}[thbp]
\begin{center}
\includegraphics[width=\columnwidth,clip]{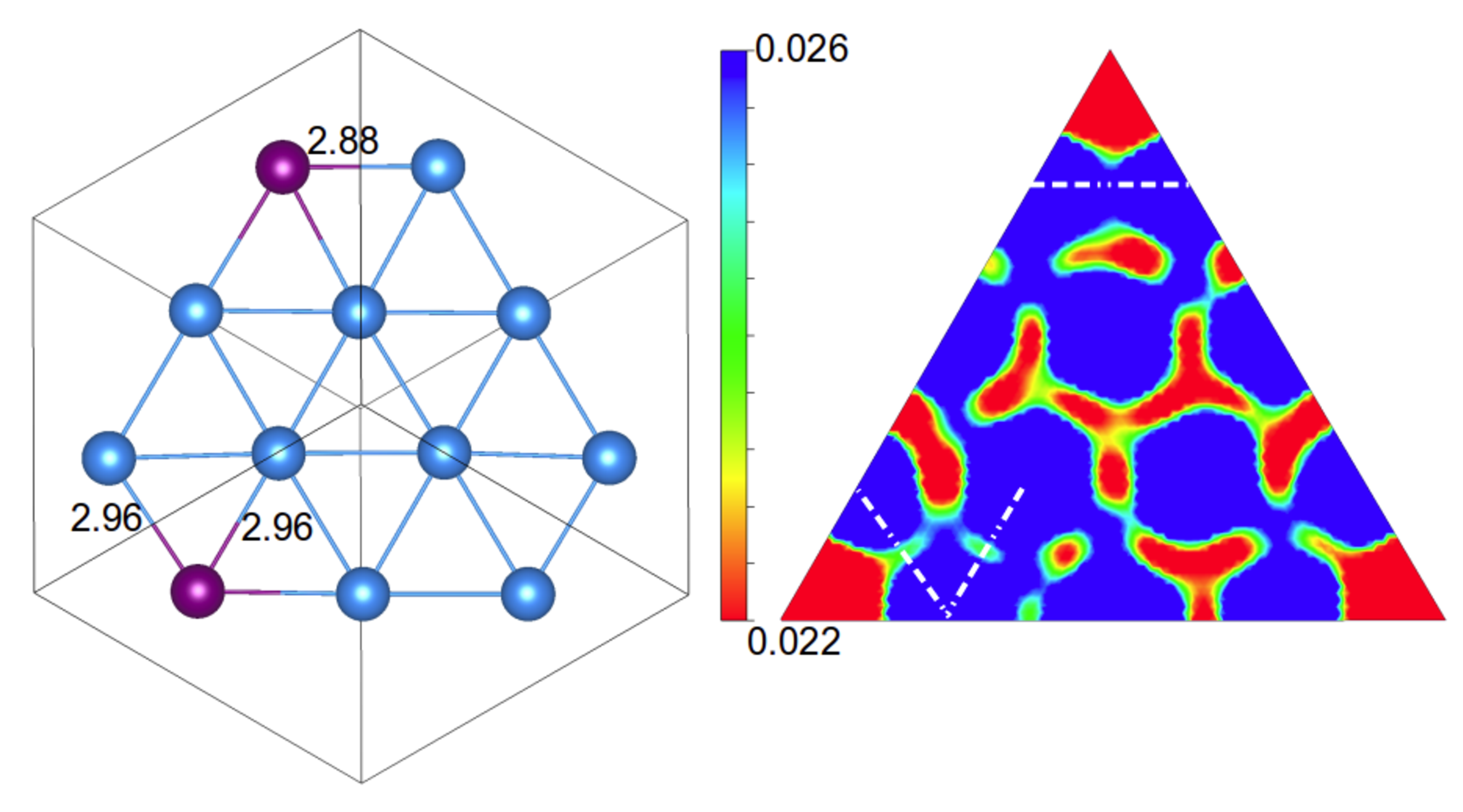}
\caption{(Color online).
Two-dimensional slice of the charge density at the (111) plane containing the V$_{Fe}$ in Fe$_{0.97}$O (right),
showing on the left the reference atomic positions. Representative interatomic distances (in $\AA$) between Fe$^{2+}$-Fe$^{3+}$ 
pairs are provided, and the corresponding CLs indicated by the dashed lines on the CD plot. \label{cargas2D-onevac}}
\end{center}
\end{figure}

In section 4 we showed that a polaronic charge distribution reminiscent of Fe$_3$O$_4$ already emerges in
the dilute limit represented by Fe$_{0.97}$O. Figure \ref{DOS-1vac}b identifies the existence of charge accumulation 
along certain shortened CLs between Fe$^{2+}_B$-Fe$^{3+}_B$ neighbors. 
An alternative view of this effect is provided in figure \ref{cargas2D-onevac},
where a two-dimensional cut of the charge density (CD) at the (111) plane containing the vacancy is presented.
In order to have a reference of the atomic centres, a sketch of the corresponding atomic positions is also shown,
with vacancies lying at the corners of the triangle. The full CD spans a range between 1.285 $e/\AA^3$
and 0.001 $e/\AA^3$, while the range of interest for interatomic charge modulations at the (111) Fe
planes is much reduced, as indicated by the color scale of the figure \cite{supplement}.
The maximum CD corresponds to the tails of the core charge, that decay very fast away from the 
atomic centres toward the interstitial regions. Usually there is a depression of the CD at the middle of the CL
between Fe neighbors. As shown by the dashed lines in the figure, it tends to be less pronounced around the 
Fe$^{3+}_B$ sites, and disappears only at the shortened Fe$^{3+}_B$-Fe$^{2+}_B$ CLs corresponding to a 
d(Fe-Fe)=2.88 $\AA$, formed by overlapping t$_{2g}$ Fe$^{2+}_B$ and Fe$^{3+}_B$ states directed along the CL.

\begin{figure}[thbp]
\begin{center}
\includegraphics[width=\columnwidth,clip]{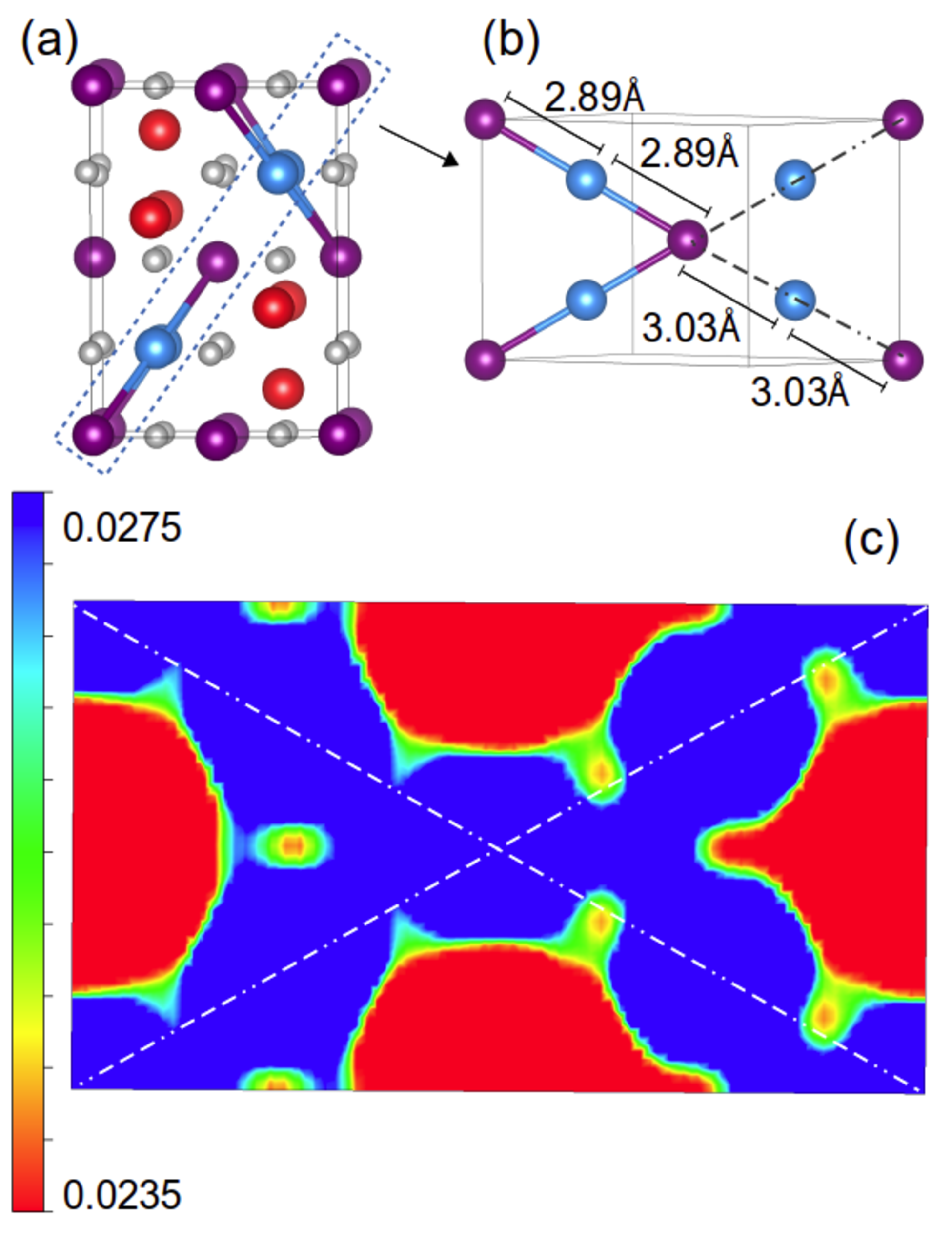}
\caption{(Color online).
(a) Sketch of a $P2/c$ unit cell of Fe$_3$O$_4$ identifying the trimerons.
The legend of figure \protect\ref{111-distortion} has been used.
(b) Top view of a plane containing two trimerons.
(c) Charge density at the plane in (b). The dashed lines follow the rows defined by the
Fe$_B$ atoms.}
\label{magnetite}
\end{center}
\end{figure}

The similitude to magnetite can be appreciated in figure \ref{magnetite}, that shows the distribution of trimerons in
a $P2/c$ unit cell of Fe$_3$O$_4$ together with the CD at a plane containing them. Notice that in magnetite the
Fe$_B$ sublattice is ferromagnetic, and antiferromagnetically coupled to the Fe$_A$ sublattice.
As explained in section 4, trimerons are linear chains of Fe$^{3+}_B$-Fe$^{2+}_B$-Fe$^{3+}_B$ n.n. with
shortened interatomic distances (2.89 $\AA$ vs. 3.03 $\AA$) \cite{Bernal-2014}. It is evident from the
figure that the trimerons inhibit the depression of the CD at the middle of the CL between adjacent 
Fe$^{3+}_B$-Fe$^{2+}_B$ sites. 
The existence of two different Fe$_A$ and Fe$_B$ sublattices modifies the charge transfer with respect to w\"ustite,
shifting the CD scale to slightly higher values.

\begin{figure}[thbp]
\begin{center}
\includegraphics[width=\columnwidth,clip]{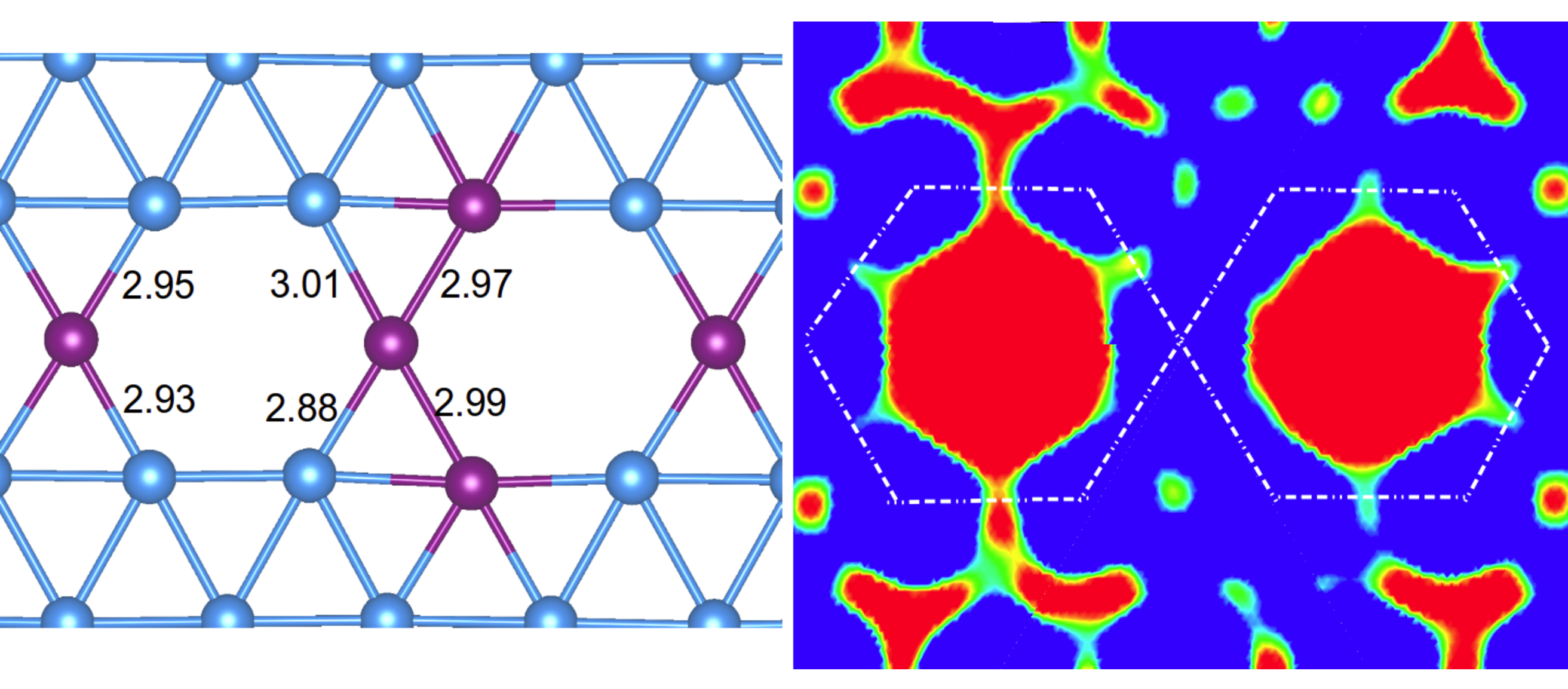}
\caption{(Color online).
Same as figure \protect\ref{cargas2D-onevac} for case C1 of Fe$_{0.906}$O, at the (111) Fe plane where the 
shortest d(Fe-Fe) can be found. \label{carga-3vac}}
\end{center}
\end{figure}

In the case of 3 isolated V$_{Fe}$ considered in section 5, the higher number of V$_{Fe}$ makes the Fe$^{3+}_B$ 
sites to be sometimes n.n. Charge accumulation is always observed in the interstitial region between adjacent 
Fe$^{3+}_B$, as evidenced in figure \ref{carga-3vac}.
But these regions of large CD lack of any directionality in the orbital order and are not coupled
to short d(Fe-Fe). Their existence, on the other hand, tends to inhibit the polaronic charge sharing between
Fe$^{3+}_B$-Fe$^{2+}_B$ neighbors. In the supercell there is only one Fe$^{3+}_B$-Fe$^{2+}_B$ distance shortened 
to 2.89 $\AA$ where polarons are formed, indicated in figure \ref{carga-3vac}.
This proves the crucial role of the long-range distribution of Fe$^{3+}_B$ in the lattice 
for the emergence of polaronic charge sharing.

\begin{figure}[thbp]
\begin{center}
\includegraphics[width=\columnwidth,clip]{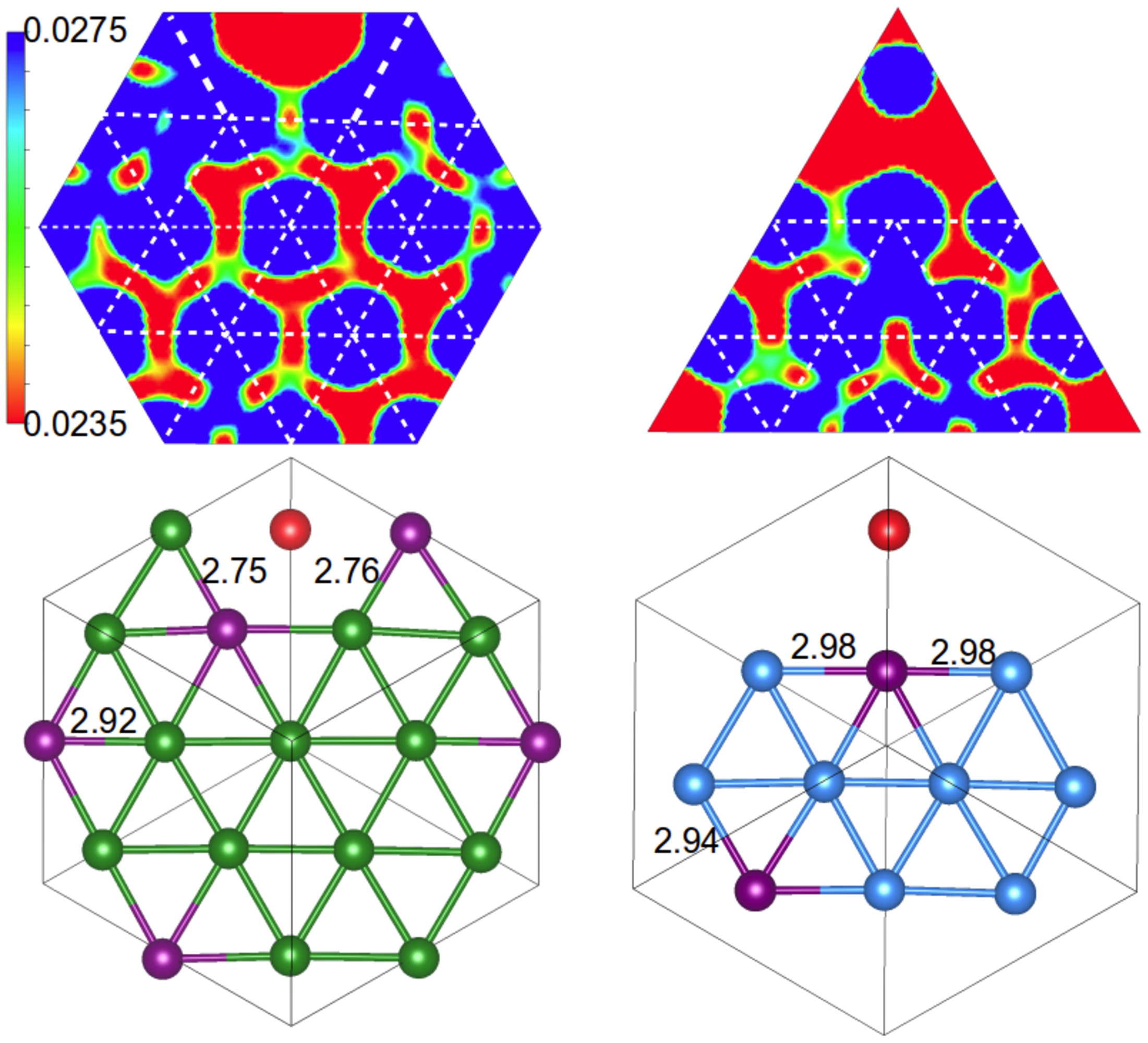}
\caption{(Color online).
CD around a 4:1 defect cluster at the two different types of (111) planes in Fe$_{0.906}$O (top).
The reference atomic positions are indicated by dashed lines on top of the CD and at the lower panel,
that also shows the projection of the Fe$_A$ position. Representative values of the interatomic Fe$^{2+}$-Fe$^{3+}$ 
distances are provided (in $\text{\AA}$).}
\label{4x1cargas}
\end{center}
\end{figure}

The similitude to magnetite is recovered at 4:1 clusters, as shown in figure \ref{4x1cargas}. 
The Fe$_A$ atom promotes a distribution where no Fe$^{3+}_B$ are first n.n.
On the other hand, it is almost coplanar along $c$ to the plane with more V$_{Fe}$, filling the void created by 
them, and inhibiting the shrinkage of the irregular blue hexagon formed by the n.n. to the vacancies
(see figure \ref{layer4x1}).
Oppositely, at the plane with one vacancy the entire hexagon formed by the Fe$_B$ 
around it shrinks, the shortest sides corresponding to Fe$^{3+}_B$-Fe$^{2+}_B$ CLs. 
These reduced d(Fe-Fe) are again accompanied by a polaronic charge distribution, 
as evidenced by the large CD (thick dashed lines in figure \ref{4x1cargas}) and the orbital character of the
corresponding Fe$^{2+}_B$ states.
The fact that there are not any d(Fe-Fe) below the threshold 2.89 $\AA$ between adjacent layers of opposite
spin polarization supports the close relation between these reduced interatomic distances and the formation
of polarons.

As the distribution of the Fe$^{3+}_B$ sites is ultimately dictated by the O positions, there is a clear 
correlation between the emergence of polaronic charge sharing and the structure
of the O fcc sublattice. Also, the influence of the Fe$_A$ sites in the distribution of the V$_{Fe}$ has an
important role to optimize the conditions for enhanced charge sharing.
A distorted O sublattice with Fe atoms at tetrahedral sites also exist in magnetite \cite{Bernal-2014}, 
where furthermore there is more freedom to form polarons along any direction of the three-dimensional space 
due to the parallel alignment of the spins of the octahedral Fe atoms. 
In essence, this links unequivocally the long-range CO of magnetite to the short-range correlations,
explaining the origin of diffuse scattering measurements that identified this connection and related the Fermi 
surface topology to the local CO \cite{Piekarz-2014}.

\section{Conclusions}

Fe$_{1-x}$O shows singular electronic properties that distinguish it from the rest of TMO
and approach to the scenario defined by Fe$_3$O$_4$. The related physical phenomena emerge
in reduced regions localized around the defects, converting Fe$_{1-x}$O in a unique material
to explore the complex interplay of lattice, charge and spin degrees of freedom involved in the
Verwey transition.

The creation of a Fe vacancy in FeO introduces charge disproportionation in the octahedral Fe sublattice,
leading to two Fe$^{3+}_B$ at n.n. positions of the vacancy site.
These Fe$^{3+}_B$ are characterized by enhanced magnetic moments and shorter bond lengths to O, 
as compared to the Fe$^{2+}_B$ of stoichiometric FeO.
They are also responsible for the closing of the insulating gap from 2 eV to 
below 1 eV. 
Their ordering in the lattice is governed by the charge balance, causing uncompensated magnetic moments
that create a net magnetization in Fe$_{1-x}$O.

In general, the Fe atoms tend to fill the voids created by the V$_{Fe}$, resulting in the contraction
of the cell volume and the reduction of the rhombohedral distortion of FeO. Regardless of the 
distribution of the V$_{Fe}$, all alterations of the structural, electronic and magnetic properties are
localized around the defects, and far from them the features of the stoichiometric oxide tend to be recovered.
Formation of compact defect clusters is clearly favored, probably linked to a more effective compensation
of the global charge.
However, the fast quenching process applied to obtain Fe$_{1-x}$O at ambient conditions may produce samples 
where different configurations coexist. 
The most relevant features introduced by the 4:1 clusters are a high degree of overall order,
due to the enhanced localization of the defects, a larger local magnetization and the existence
of more favorable conditions to form polarons.

In analogy to magnetite, the emergence of polaronic charge sharing is linked to the anomalous shortening of 
certain interatomic Fe$^{3+}_B$-Fe$^{2+}_B$ distances below a threshold value of 2.89 $\AA$.
The formation of polarons is conditioned by the distribution of 
the V$_{Fe}$ and the Fe$^{3+}_B$ sites, which in turn depends on the internal structure
of the O sublattice. This connects the local charge order to the long-range one, a phenomenon also observed
in magnetite \cite{Piekarz-2014}, thus allowing to understand the origin of the relation between
short-range correlations and charge order across the Verwey transition.

The influence of the magnetic interactions is at the same time complex and subtle.
Together with the reduction of the magnetically induced rhombohedral distortion,
the limited effect in the energy balance of local alterations of the magnetization around 
the defects supports a secondary role of the magnetism in the stability of the system.
However, the existence of a distortion at the O fcc sublattice that requires the presence of magnetism
is preserved when V$_{Fe}$ are introduced. Furthermore, this distortion plays a crucial role in the
emergence of polaronic charge distributions, which are favored under the most stable defect configurations.
The importance of the magnetic interactions is higher when tetrahedral Fe$_A$ sites are 
occupied, making plausible that the magnetic energy balance becomes crucial to determine
the stability of large defect structures based on spinel-like clusters.

\section{Acknowledgments}

This work has been financed by the Spanish Ministry of Science under contract MAT2012-38045-C04-04.
I.B. acknowledges financial support from the JAE program of the CSIC.

\providecommand{\noopsort}[1]{}\providecommand{\singleletter}[1]{#1}%

\end{document}